%% file: main.tex
\documentclass{article}

\usepackage{microtype}
\usepackage{graphicx}
\usepackage{subfigure}
\usepackage{booktabs} %
\usepackage{amsmath}
\usepackage{amssymb}
\usepackage{esvect}
\usepackage{xspace}
\usepackage{enumitem}
\usepackage{adjustbox}
\usepackage{xcolor}
\usepackage{textcomp}
\usepackage{hyperref}
\usepackage{nicefrac}

\newcommand{\pluseq}{\mathrel{+}=}

\newcommand{\sys}{Punica\xspace}
\newcommand{\mytitle}{\sys: Multi-Tenant LoRA Serving}

\usepackage{hyperref}

\def\Snospace~{\S{}}

\usepackage[accepted]{mlsys2024-arxiv}

\mlsystitlerunning{\mytitle}

\begin{document}

\twocolumn[
\mlsystitle{\mytitle}

\mlsyssetsymbol{equal}{*}

\begin{mlsysauthorlist}
\mlsysauthor{Lequn Chen}{equal,uw}
\mlsysauthor{Zihao Ye}{equal,uw}
\mlsysauthor{Yongji Wu}{duke}
\mlsysauthor{Danyang Zhuo}{duke}
\mlsysauthor{Luis Ceze}{uw}
\mlsysauthor{Arvind Krishnamurthy}{uw}
\end{mlsysauthorlist}

\mlsysaffiliation{uw}{University of Washington}
\mlsysaffiliation{duke}{Duke University}

\mlsyscorrespondingauthor{Lequn Chen}{lqchen@cs.washington.edu}

\mlsyskeywords{Large Language Model, Batching, Serving, Fine-tuning, LoRA, LLM}

\vskip 0.3in

\begin{abstract}
\input{tex/00_abstract}

\end{abstract}
]

\printAffiliations{\mlsysEqualContribution} %

\input{tex/10_intro}
\input{tex/20_background}

\input{tex/30_design}

\input{tex/40_cuda_kernel}
\input{tex/50_scheduling}

\input{tex/60_impl}
\input{tex/70_eval}

\input{tex/90_related}
\input{tex/99_conclusion}

\bibliography{reference}
\bibliographystyle{mlsys2024}

\end{document}

%% file: tex/00_abstract.tex
Low-rank adaptation (LoRA) has become an important and popular method to adapt pre-trained models to specific domains.
We present Punica, a system to serve multiple LoRA models in a shared GPU cluster. Punica contains a new CUDA kernel design that allows batching of GPU operations for different LoRA models. This allows a GPU to hold only a single copy of the underlying pre-trained model when serving multiple, different LoRA models, significantly enhancing GPU efficiency in terms of both memory and computation. Our scheduler consolidates multi-tenant LoRA serving workloads in a shared GPU cluster. With a fixed-sized GPU cluster, our evaluations show that Punica achieves 12x higher throughput in serving multiple LoRA models compared to state-of-the-art LLM serving systems while only adding 2ms latency per token.
Punica is open source at \url{https://github.com/punica-ai/punica}.

%% file: tex/10_intro.tex
\section{Introduction}

Low-rank adaptation (LoRA)~\cite{lora} is becoming increasingly popular in specializing pre-trained large language models (LLMs) to domain-specific tasks with minimal training data. 
LoRA retains the weights of the pre-trained model and introduces trainable rank decomposition matrices to each layer of the Transformer architecture, significantly reducing the number of trainable parameters and allowing tenants to train different LoRA models at a low cost.
LoRA has been integrated into many popular fine-tuning frameworks~\cite{hf-peft}.
Consequently, ML providers have to serve a large number of specialized LoRA models simultaneously for their tenants' needs.

Simply serving LoRA models as if they were independently trained from scratch wastes GPU resources. Assuming we need $k$ GPUs to serve each LoRA model, serving $n$ different LoRA models would seemingly require $k\times n$ GPUs. This straightforward approach overlooks the potential weight correlations among these LoRA models, given they originate from the same pre-trained models. 

We believe an efficient system to serve multiple, different LoRA models needs to follow three design guidelines. (G1) GPUs are expensive and scarce resources, so we need to consolidate multi-tenant LoRA serving workloads to a small number of GPUs, increasing overall GPU utilization. 
(G2) As prior works have already noticed~\cite{orca}, batching is one of the, if not the most, effective approaches to consolidate ML workloads to improve performance and GPU utilization. However, batching only works when requests are for the exact same model. We thus need to enable batching for different LoRA models. (G3) The decode stage is the predominant factor in the cost of model serving. We thus only need to focus on the decode stage performance. Other aspects of the model serving are less important, and we can apply straightforward techniques, e.g., on-demand loading of LoRA model weights.

Based on these three guidelines, we design and implement \sys, a multi-tenant serving framework for LoRA models on a shared GPU cluster. One key novelty is the design of a new CUDA kernel, \textbf{Segmented Gather Matrix-Vector Multiplication} (SGMV). SGMV allows batching GPU operations for the concurrent execution of multiple, different LoRA models. With SGMV, a GPU only needs to store a single copy of the pre-trained model in memory, significantly improving GPU efficiency in terms of both memory and computation. We pair this new CUDA kernel with a series of state-of-the-art system optimization techniques.

SGMV allows batching requests from different LoRA models, and \textit{surprisingly}, we observe negligible performance differences between batching the same LoRA models and batching different LoRA models. At the same time, the on-demand loading of LoRA models has only millisecond-level latency. This gives \sys the flexibility to consolidate user requests to a small set of GPUs without being constrained by what LoRA models are already running on the GPUs.

\sys thus schedules multi-tenant workloads in the following two ways. For a new request, \sys routes the request to a small set of active GPUs, ensuring that they reach their full capacity. Only when the existing GPUs are fully utilized, \sys will allocate additional GPU resources. For existing requests, \sys periodically migrates them for consolidation. This allows freeing up GPU resources that are allocated to \sys.

We evaluate LoRA models that are adapted from Llama2 7B, 13B, and 70B models~\cite{llama2} on NVIDIA A100 GPU clusters. Given the same amount of GPU resources, \sys achieves 12x higher throughput
compared to state-of-the-art LLM serving systems while only adding 2ms latency per token.

This paper makes the following contributions:
\begin{itemize}
    \item We identify the opportunity of batch processing requests of multiple, different LoRA models.
    \item We design and implement an efficient CUDA kernel for running multiple LoRA models concurrently.
    \item We develop new scheduling mechanisms to consolidate multi-tenant LoRA workloads.
\end{itemize}

%% file: tex/20_background.tex
\section{Background}

We first present the text generation process for transformer models. We then describe Low-Rank Adaptation (LoRA) of transformer models.

\subsection{Transformer and Text Generation}
Transformer-based LLMs operate on a sequence of tokens.
A token is roughly \textthreequarters{} of an English word.
An LLM's operation consists of two stages.
The \textit{prefill} stage accepts a user prompt and generates a subsequent token and a Key-Value cache (KvCache).
The \textit{decode} stage accepts a token and the KvCache, and it then generates one more token and appends a column in the KvCache.
The decode stage is an iterative process.
The generated token then becomes the input for the next step.
This process ends when the end-of-sequence token is generated.

\begin{figure}[t]
    \centering
    \includegraphics[width=\linewidth]{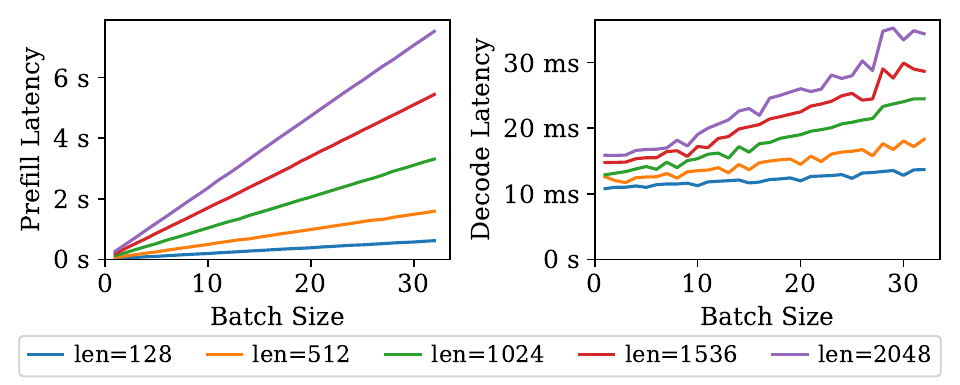}
    \caption{Batching effects in Prefill stage and in Decode stage}
    \label{fig:prefill-vs-decode-latency}
\end{figure}

A transformer block contains a self-attention layer and a multilayer perceptron (MLP). Let us assume that the length of the prompt is $s$ and the attention head dimension is $d$. For the prefill stage, the computation of the self-attention layer is $(s, d) \times (d, s) \times (s, d)$, and the MLP computation is $(s, h) \times (h, h)$. For a decode step, assuming $s$ represents the past sequence length, the computation of the self-attention layer is $(1, d) \times (d, s+1) \times (s+1, d)$ and the MLP computation is $(1, h) \times (h, h)$. The decode stage has low GPU utilization because the input is a single vector.

\autoref{fig:prefill-vs-decode-latency} shows the latency for the prefill stage and the decode stage for different batch sizes. The computation capability of the GPU is fully utilized during the prefill stage. Prefill latency is proportional to batch size. However, this is not the case for the decode stage. Increasing the batch size from 1 to 32, the decode step latency increases from 11ms to 13ms for short sequences, and from 17ms to 34ms for longer sequences. This means that batching can improve GPU utilization significantly for the decode stage. Orca~\cite{orca} leveraged this opportunity to build an efficient LLM serving system. This type of batching is especially important because the decode stage predominately determines the serving latency for long output length responses.

\subsection{Low-Rank Adaptation (LoRA)}
\label{sec:lora}

Fine-tuning allows a pre-trained model to adapt to a new domain or a new task or be improved with new training data. However, because LLMs are large, fine-tuning all the model parameters is resource-intensive.

Low-Rank Adaptation (LoRA)~\cite{lora} significantly reduces the number of parameters needed to be trained during fine-tuning. The key observation is that the weight difference between the pre-trained model and the model after fine-tuning has a low rank.
This weight difference can thus be represented as the product of two small and dense matrices. LoRA fine-tuning then becomes similar to training 
a small, dense neural network. Formally, let's consider the weights of the pre-trained model to be $W \in \mathbb{R}^{h_1 \times h_2}$. LoRA fine-tuning trains two matrices $A \in \mathbb{R}^{h_1 \times r}$ and $B \in \mathbb{R}^{r \times h_2}$, where $r$ is the LoRA Rank. $W+AB$ is the new weight for the fine-tuned model. LoRA rank is usually much smaller than the original dimension (e.g., 16 instead of 4096). In addition to fast fine-tuning, LoRA has very low storage and memory overheads. Each fine-tuned model only adds 0.1\% to 1\% of the model weight.
LoRA is usually applied to all dense projections in the transformer layer~\cite{qlora}, including the Query-Key-Value-Output projections in the attention mechanism and the MLP. Note that the self-attention operation itself does not contain any weight.

\paragraph{How to serve multi-tenant LoRA models efficiently on a shared GPU cluster?}
LoRA provides an efficient algorithm to fine-tune LLMs. Now the question is: how to serve those LoRA models efficiently? One approach is to regard each LoRA model as an independent model and use traditional LLM serving systems (e.g., vLLM). However, this neglects the weight sharing among different LoRA models that can be used to significantly improve GPU efficiency. Further, if we treat each LoRA model as an independent one, model loading time can be a substantial performance bottleneck when bootstrapping model serving.
Even if we share the backbone model across the LoRA models, it remains a question as to how to batch compute the LoRA addon efficiently.

%% file: tex/30_design.tex
\section{\sys Overview}

\begin{figure}[t]
    \centering
    \includegraphics[width=\linewidth]{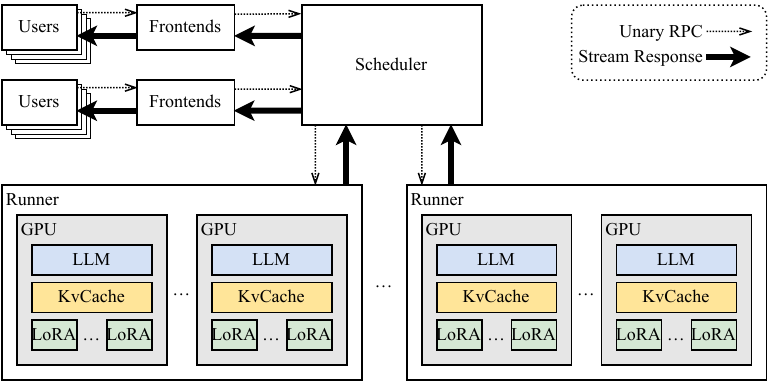}
    \caption{The system architecture of \sys.}
    \label{fig:punica-diagram}
\end{figure}

We design \sys as a multi-tenant system that manages a cluster of GPUs to serve multiple LoRA models with shared pre-trained backbone models. \autoref{fig:punica-diagram} shows the system architecture of \sys. Like other model serving systems, \sys has frontend servers that expose RESTful API to end-users and forward users' serving requests to the \sys scheduler. A user request contains the identifier of the LoRA model and a prompt.
The scheduler dispatches requests to the GPUs.
Each GPU server starts a runner, which communicates with the scheduler and controls the execution of all the GPUs.
As GPUs generate new tokens, new tokens are streamed from the runners to the scheduler, to the frontends, and finally to the end-users.

In \sys, each GPU loads the backbone pre-trained large language model.
A large fraction of GPU memory is reserved for KvCache.
Only the LoRA components of models are swapped in from remote storage when needed. Note that this design allows fast cold-start for model serving. Because the pre-trained model is already loaded into the GPU memory, \sys only needs to load matrices A and B for a new LoRA model.

\sys needs to address two key research challenges. The first challenge is how to run multiple LoRA models efficiently on a GPU. Because requests have to be served by different LoRA models, each request has to go through a different GPU computation.
We use the existing matrix multiplication for the backbone computation.
And we present a new CUDA kernel for adding the LoRA addons to the backbone computation in a batched manner.
We call this kernel \emph{Segmented Gather Matrix-Vector Multiplication} (SGMV). 
SGMV parallelizes the feature-weight multiplication of different requests in the batch and groups requests corresponding to the same LoRA model to increase the operational intensity of the kernel and use GPU Tensor Cores units for acceleration.

The second challenge is how to design an efficient system on top of SGMV for multi-tenant LoRA model serving. 
Our goal here is to consolidate multi-tenant workloads to the smallest set of GPUs possible, occupying the least amount of GPU resources. \sys schedules user requests to active GPUs, which already serve or train LoRA models. This is feasible in \sys, because with SGMV adding the batch size, even if for different LoRA models, improves the GPU utilization. For old requests, \sys migrates them periodically to consolidate the workloads, thereby freeing up GPU resources.

Next, we describe the details of \sys's CUDA kernel and other design details of \sys.

%% file: tex/40_cuda_kernel.tex
\section{Segmented Gather Matrix-Vector Multiplication}

\begin{figure}[t]
    \centering
    \includegraphics[width=0.7\linewidth]{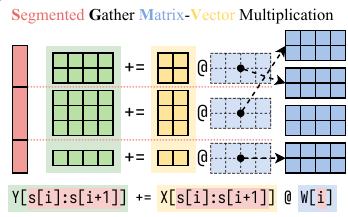}
    \caption{Semantics of SGMV.}
    \label{fig:sgmv-semantics}
\end{figure}

When a LoRA model has multiple inputs in the batch, we can further batch them together.
We group inputs to the same LoRA model consecutively.
Denote $n$ as the number of LoRA models in a batch.
Denote sequence $s_i$ as the last element index for $i$-th model within the batch.
In particular, $s_0=0$ and $s_n$ is the batch size.
Input $\left\{ \vv{x_i}  \mid i \in [1,s_n] \right\}$ is then partitioned as
$\left\{ \left\{\vv{x_j}  \mid j \in (s_{n-1}, s_n]\right\} \mid i \in [1,n] \right\}$.
The dense projection output can then be written as:
\begin{equation*}
\resizebox{\linewidth}{!}{
$
\begin{pmatrix}
\begin{pmatrix}
\vv{y_{1}} \\
\vdots \\
\vv{y_{s_1}}
\end{pmatrix} \\
\vdots \\
\begin{pmatrix}
\vv{y_{s_{n-1}+1}} \\
\vdots \\
\vv{y_{s_n}}
\end{pmatrix} \\
\end{pmatrix}
:=
\begin{pmatrix}
\begin{pmatrix}
\vv{x_{1}} \\
\vdots \\
\vv{x_{s_1}}
\end{pmatrix} \\
\vdots \\
\begin{pmatrix}
\vv{x_{s_{n-1}+1}} \\
\vdots \\
\vv{x_{s_n}}
\end{pmatrix} \\
\end{pmatrix}
W
+
\begin{pmatrix}
\begin{pmatrix}
\vv{x_{1}} \\
\vdots \\
\vv{x_{s_1}}
\end{pmatrix} A_1B_1 \\
\vdots \\
\begin{pmatrix}
\vv{x_{s_{n-1}+1}} \\
\vdots \\
\vv{x_{s_n}}
\end{pmatrix} A_nB_n \\
\end{pmatrix}
$
}
\end{equation*}

The left-hand-side multiplication is the computation for the backbone model, which is batched through regular GEMM.
We need a fast kernel to compute the right-hand-side LoRA addon.
Note that operator $\vv{y} \pluseq \vv{x}AB$ can be separated as two launches of the same kernel:
Initialize $\vv{v} := \vv{0}$. Then we run $\vv{v} \pluseq \vv{x}A$ and follow by $\vv{y} \pluseq \vv{v}B$.

We name this operator SGMV, Segmented Gather Matrix-Vector multiplication.
Figure~\ref{fig:sgmv-semantics} illustrates the semantics of SGMV.

\paragraph{CUDA Kernel Schedule}
We classify SGMV operator into two categories, SGMV-shrink and SGMV-expand, based on their input and output feature dimensions. The first operator in the LoRA module: $\vv{v} = \vv{x}A$ is SGMV-shrink because it shrinks a high-dimensional input feature to low-rank output. The second operator $\vv{y} = \vv{v}B$ is SGMV-expand as it expands the low-rank input feature to a high-dimensional output feature.

\begin{figure}
    \centering
    \includegraphics[width=\linewidth]{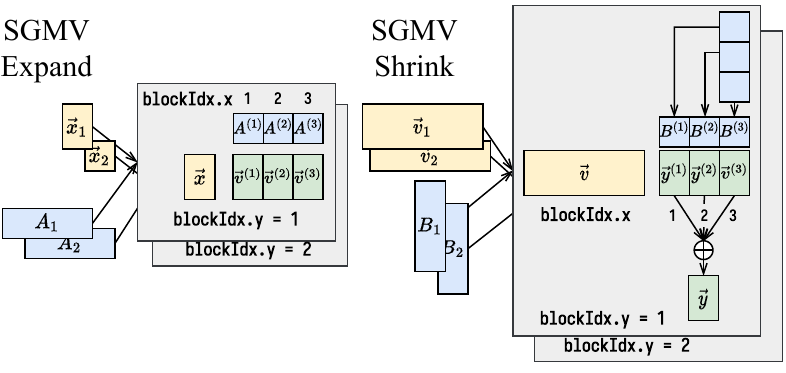}
    \caption{Scheduling of SGMV expand/shrink kernels}
    \label{fig:sgmv-kernel}
\end{figure}

\autoref{fig:sgmv-kernel} shows how we schedule the SGMV kernel in these two cases:
For both kernels, we bind the LoRA index to \textsc{blockIdx.y} in CUDA. Then, the computation on each \textsc{blockIdx.y} is a matrix multiplication between features and a specific LoRA weight. We designed different schedules for matrix multiplication under expand and shrink settings: for the expand kernel, we split $A$ on the output feature dimension $A = \begin{bmatrix}A^{(1)} & \cdots & A^{(n)}\end{bmatrix}$ and dispatch different $\vv{v}^{(i)} = \vv{x} A^{(i)}$ to different threadblocks in GPU, and the concatenation $\vv{v}^{(i)}$ on different threadblocks forms the final result $\vv{v} = \begin{bmatrix}\vv{v}^{(1)} & \cdots & \vv{v}^{(n)}\end{bmatrix}$; for the shrink kernel, the output dimension is too thin and we adopt the Split-K strategy ~\cite{cutlass} to increase parallelism: we split $B$ on the input feature dimension $B = \begin{bmatrix}B^{(1)} \\ \cdots \\ B^{(k)}\end{bmatrix}$. We dispatch different $\vv{y}^{(i)} = \vv{v} B^{(i)}$ to different threadblocks in GPU, after the partial sum $\{\vv{y}^{(i)}\}$ computation on all threadblocks are finished, we perform a grid synchronization followed by a cross threadblock reduction $\vv{y} = \sum_{i=1}^{k} \vv{y}^{(i)} $ to aggregate the partial results. We use GPU Tensor Cores to accelerate matrix multiplication for both kernels.

In the case that each request has a distinct LoRA index, the computation corresponding to each LoRA index degrades to matrix-vector multiplication, which is totally IO-bound. We designed a specific schedule for this case that maximizes memory bandwidth utilization and does not use Tensor Cores because of the low operational intensity of the operator.

%% file: tex/50_scheduling.tex
\section{\sys in Detail}

\sys schedules new user requests at a per-request level and migrates old requests between GPUs at a per-iteration level. The scheduler adds requests to a GPU or cancels a working request from a GPU. Each GPU batches all requests in its working set for LLM invocation.
GPU runs the Prefill steps and Decode steps continuously.
When a request reaches the stopping condition (end-of-sequence token or length limit), the GPU removes the request from the batch and notifies the scheduler about the stopping.

We run batch requests of prefill and decode stages in a single model invocation.
To minimize latency penalty, we limit the prefill batch size to 1 for each batch.
The single prefill and the batch of decodes invoke two separate CUDA kernels for the self-attention operation.
All other operations, including dense projection and LoRA addon, treat all tokens in the prefill stage and decode stage as a single batch input.
In this way, we increase the batch efficiency of dense projection and LoRA addon.

\subsection{Scheduling new requests}

\sys scheduler has a global view of the state of all the GPUs. In particular, for each GPU, \sys maintains the working set of requests, which is the batch input of LLM invocation. As new requests are added to the working set and as the decode steps unfold, the KvCache consumes more and more GPU memory. Therefore, \sys also continuously tracks the memory space available for KvCache on each GPU.

\sys schedules a new request to the GPU that currently has the largest working set of requests (i.e., the LLM invocation batch size) while satisfying the following constraints:
(1) It has not yet reached the max batch size limit.
(2) It has enough memory for the new request's KvCache.
When there are multiple candidates, the one that has the highest GPU UUID gets the new request.
When all GPUs are fully occupied (i.e., have reached the maximum batch size or have insufficient memory), the request is queued.
When some GPUs become available in the future, queued requests are scheduled in a first-come-first-serve (FCFS) manner.

The max batch size limit balances the cluster throughput and the per-token latency. Oversized batches greatly slow down latency while providing marginal throughput gains. We profile A100 GPUs and decide to set the maximum batch size to 32.

The GPU selection logic emphasizes cluster throughput within the latency sweet spot. Our scheduling approach has the following attributes: a busy GPU is likely to stay busy as more requests will be assigned to it, a lightly loaded GPU is likely to lower its load as requests terminate, and an idle GPU is likely to stay idle. As a consequence, our scheduler maintains peak throughput and consolidates GPU usage based on the current overall system load. This allows easier decisions to scale up/down the GPU cluster. In a cloud setting where \sys can allocate and deallocate GPU servers, we perform the following cluster allocations: (1) If no lightly loaded GPU exists in the cluster, \sys should request more GPUs. (2) \sys can return the GPU resources for GPU servers with no load.

\subsection{On-demand model loading}

Weight sharing between the LoRA model and the underlying pre-trained model makes model loading fast. The size of the LoRA model (which is matrices $A$ and $B$ in \autoref{sec:lora}) is only 1\% of the underlying pre-trained model.

Loading a LoRA model from the main memory to the GPU memory is merely an asynchronous host-to-device memory copy.
The latency is bounded by the PCIe bandwidth.
On PCIe Gen4 x16, it takes around 50\textmu{}s to load a layer and 2ms to load the entire model.
Since the memory copy and the GPU computation can overlap, it is feasible to implement sophisticated layer-by-layer or even matrix-by-matrix loading to minimize the model loading delay.

However, notice that each decode step takes around 30ms to complete, and each request might need thousands of decode steps.
We opt to use a simpler yet equivalently efficient method.
When a request is newly added to a GPU, if its LoRA model is not already loaded, we issue an asynchronous memory copy to load the LoRA weight, and let the GPU continue running other inputs in the batch.
By the end of the model execution, the weight already finished loading.
Then, the new request is able to join the batch naturally.

\subsection{Request migration}

\begin{figure}[t]
    \centering
    \includegraphics[width=\linewidth]{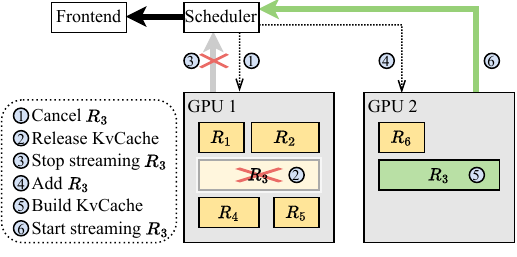}
    \caption{Request migration procedure for Request $R_3$.}
    \label{fig:punica-migratin-diagram}
\end{figure}

As each request generates more tokens, their KvCache occupies more GPU memory.
When a GPU runs out of space for KvCache, it migrates some requests to other GPUs.
The request migration consists of two steps --- evict and add.
The scheduler evicts the newest request from the GPU. This preserves the FCFS semantics.
The scheduling for the evicted request is the same as adding a new request.

\sys scheduler supports canceling requests. Cancellation is straightforward: remove the request from both the GPU and the scheduler states. A typical scenario for cancellation is user disconnection. More importantly, request cancellation as a scheduling primitive enables request migration.

\autoref{fig:punica-migratin-diagram} shows the workflow to migrate a request, $R_3$, from GPU 1 to GPU 2.
The scheduler first sends the cancellation of the request to GPU 1.
After GPU 1 finishes running the previous batch, it picks up the cancelation and releases KvCache.
GPU 1 also omits the $R_3$'s new token generated in the previous batch.
Immediately after sending the cancellation to GPU 1, the scheduler adds $R_3$ to GPU 2.
GPU 2 runs a prefill step on $R_3$'s original prompt plus all previously generated tokens.
This reinstablishes the $R_3$'s KvCache on GPU 2.
GPU 2 then starts to stream $R_3$'s new tokens to the scheduler.

We opt for recomputation instead of moving the KvCache for its simplicity.
PagedAttention~\cite{PagedAttention} has shown that recomputation's latency is equal to or better than moving the KvCache in most cases, which agrees with our observation.

\subsection{Memory layout for KvCache}

\sys uses a separable KvCache layout, which is important for text generation batching throughput.
The HuggingFace Transformers library's KvCache layout consists of complicated nested lists of tensors, which can conceptually be viewed as the following shape:
\[
[L, 2, B, N, S, D]
\]
where $L$ is the number of layers, $2$ is for Key and Value projection, $B$ is batch size, $N$ is the number of heads, $S$ is the sequence length, and $D$ is the head dimension.
In each decode step, HuggingFace Transformers concatenates one tensor along the sequence length dimension.
The concatenation is inefficient as it needs to read the whole KvCache and write a new copy,
whereas the new tensor is only $1/S$ of the KvCache.

The bigger problem with the HuggingFace Transformer's approach is that the batching dimension is not the outermost dimension, which means that requests in the batch are hard to separate in the KvCache.
Under this restriction, requests that enter the batch together need to remain together during all decode steps \emph{until all requests meet their own stopping condition.}

\begin{figure}[t]
    \centering
    \includegraphics[width=0.9\linewidth]{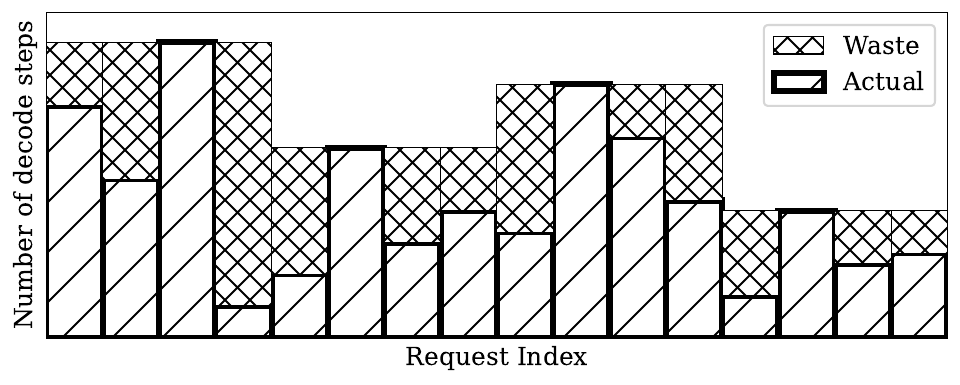}
    \caption{Inseparable KvCache adds wasted decode steps.}
    \label{fig:sketch-inseparable-kv}
\end{figure}

\autoref{fig:sketch-inseparable-kv} is an illustrative figure that explains the problem.
In the figure, consecutive 4 requests are batched together.
The striped bar represents the number of decode steps that each request actually needs to reach its stopping condition.
Due to inseparable KvCache, shorter requests in the batch run additional decode steps, which is essentially wasted computation.
FasterTransformer ~\cite{FasterTransformer} and DeepSpeed ~\cite{DeepSpeed-Inference} also suffer from similar problems.

Instead, our KvCache layout is
\[\textstyle
\left[\sum_i \left\lceil \nicefrac{S_i}{P} \right\rceil, L, 2, N, P, D\right]
\]
where $S_i$ is the length of the sequence $i$ and $P$ is the page size. We use paged KvCache~\cite{PagedAttention} to minimize memory fragmentation. We put the batching dimension on the outmost to enable continuous batching.

%% file: tex/60_impl.tex
\section{Implementation}

\sys implementations consist of two parts: a Python library on top of PyTorch that runs large language models on a single GPU and other system components to support model serving across a GPU cluster.

\paragraph{Python Library}
We expose our CUDA kernels as a PyTorch Extension using PyBind11.
Llama model implementation is adapted from the HuggingFace Transformers library.
We use FlashInfer~\cite{flashinfer} open source project for fast and memory-efficient computation of self-attention.
Besides fusing the computation of $\mathrm{softmax}(QK^T)V$ like FlashAttention~\cite{FlashAttention} does, FlashInfer further supports batch decoding without padding.
Similar to PagedAttention~\cite{PagedAttention}, FlashInfer supports paged KvCache to minimize GPU memory fragmentation due to KvCache.
We also fuse LayerNorm, which reduces latency from 110\textmu{}s to 4\textmu{}s.

We mix new requests in the Prefill stage and existing requests in the Decode stage in one batch together.
This way, dense projections and LoRA can benefit from a bigger batch size.
For batching, we concatenate all inputs along the sequence length dimension.
We always put Prefill requests at the beginning and Decode requests at the latter part.
We then pass a \verb|BatchLen| struct to distinguish different requests.
\verb|BatchLen| contains a list of indices indicating the starting index of each Prefill request.
It also contains a number indicating the number of Decode requests.
In the self-attention layer, we pass the indices and leading input states to the BatchPrefill kernel, and we pass the trailing input states to the BatchDecode kernel.
Within a batch, we further organize the batch input order such that requests that share the same LoRA model are consecutive. The tail of Prefill requests and the head of Decode requests can share a LoRA model if possible. We then generate the segment indices for the SGMV kernel.
Before each batched model invocation, we concatenate batch inputs and construct \verb|BatchLen| and SGMV segment indices. Both \verb|BatchLen| and SGMV segment indices remain constant for the entire model invocation. This design avoids recomputation ($L$ times for \verb|BatchLen| and $7L$ times for SGMV segment indices, where $L$ is the number of layers).

\paragraph{Other system components}
We write our scheduler, frontend, and runner in Rust.
Unary RPC and streaming text chunks are both implemented via web socket.
I/Os are handled asynchronously. Runner spawns a Python subprocess for each GPU.
The subprocess is a thin warper around our PyTorch library.
The Runner main process communicates with the subprocesses using pipes.

%% file: tex/70_eval.tex
\section{Evaluation}
We evaluate \sys on two testbeds. Testbed \#1 is a single server with one NVIDIA A100 80GB GPU. Testbed \#2 consists of two NVIDIA HGX A100 40GB servers with 8 GPUs on each server.
Testbed \#1 contains a GPU with large GPU memory, which allows us to study the LoRA batching effect. Testbed \#2 is equipped with fast NvSwitch technology for us to study tensor parallelism and evaluate cluster deployment.
We use the Llama-2~\cite{llama2} model with 7B, 13B, and 70B parameters. For all experiments, we use 16 as the LoRA rank. LoRA is applied to all dense projections. We use random weights for LoRA models as the weight does not affect latency performance.

\paragraph{Workloads}
Prompt lengths and response lengths are key workload characteristics for LLM serving. We use the prompt and response length distributed from ShareGPT~\cite{sharegpt}, which consists of user-bot conversations from Internet users. We consider four types of request distribution among LoRA models.
(1) \textbf{Distinct}: each request is for a distinct LoRA model.
(2) \textbf{Uniform}: all LoRA models are equally popular. Given $n$ requests, we use $\left\lceil\sqrt{n}\right\rceil$ models.
(3) \textbf{Skewed}: model popularity follows Zipf-$\alpha$ distribution. The number of requests to the $i$-th most popular model is $\alpha$ times that of the $i+1$-th's. In our experiments, we choose $\alpha$ to be $1.5$.
(4) \textbf{Identical:} all requests are for the same LoRA model.

\paragraph{Baselines}
Since there is no well-known multi-LoRA serving system, we compare \sys against a variety of popular LLM backbone serving systems. We allow various degrees of relaxation that are in favor of baseline systems. We use HuggingFace PEFT~\cite{hf-peft} library to add LoRA weights to HuggingFace Transformers~\cite{hf-transformers} library and DeepSpeed~\cite{DeepSpeed-Inference}. We run backbone-only for FasterTransformer~\cite{FasterTransformer} and vLLM\footnote{With  \href{https://github.com/vllm-project/vllm/commit/928de46888b9b257dfa491047a7d9cd199ca585b}{commit \texttt{928de46}}}~\cite{PagedAttention} since these two systems do not support LoRA models. We omit the model switching costs for baseline systems.

\subsection{Microbenchmarks}

We use analysis and testbed evaluations to benchmark SGMV and distill the -performance implications for the LoRA operator and a single transformer layer.

\begin{figure}[t]
    \centering
    \includegraphics[width=\linewidth]{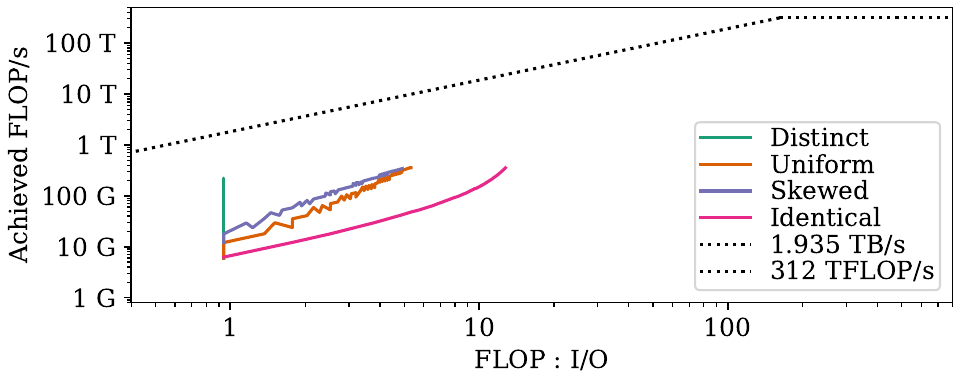}
    \caption{Roofline plot of the SGMV kernel.}
    \label{fig:sgmv-roofline}
\end{figure}

\paragraph{Roofline analysis for SGMV}
First, we use the roofline model~\cite{roofline} to understand the performance of our SGMV kernel.
The number of floating point operations (FLOP) and the number of memory I/O bytes of SGMV are calculated as:
\begin{align*}
\mathrm{FLOP}=& s_n \times h_i \times h_o \times 2 \\
\mathrm{I/O} =& [s_n \times (h_i+h_o) + n \times h_1 \times h_2] \times 2
\end{align*}
where $n$ is the number of LoRA models, $s$ is the segment indices, $s_n$ is the total number of inputs, and $h_i$ and $h_o$ are the input and output dimensions of the SGMV weight matrix.
The factor 2 in FLOP comes from multiply-add operations for matrix multiplication.
The factor 2 in I/O comes from the byte size of 16-bit floating point data type.
We use $h_i=16$, $h_o=4096$ for this case study.
We measure the latency of batch size 1 to 64 under the four different popularity distributions on Testbed \#1.

\autoref{fig:sgmv-roofline} shows the roofline model plot.
The x-axis in the roofline model is arithmetic intensity, which is defined as the ratio of FLOP to I/O.
The y-axis is the achieved throughput in terms of FLOP per second, calculated using measured latency.
The diagonal dotted line and the top dotted line represent the memory bandwidth and peak FP16 performance of the NVIDIA A100 GPU, respectively.

In the \textbf{Distinct} case, the arithmetic intensity does not change because FLOP and I/O grow at the same rate.
Since each input only utilizes a small amount of GPU compute units, increasing the batch size increases performance.
In the \textbf{Identical} case, the line goes up diagonally following the slope of memory bandwidth, which means that SGMV is bounded by memory bandwidth.
The \textbf{Uniform} case and the \textbf{Skewed} case sit in between, as a combination of both effects, the increasing degree of parallelism and the increasing arithmetic intensity.

\paragraph{LoRA operator microbenchmark}

\begin{figure*}[ht]
    \centering
    \includegraphics[width=\linewidth]{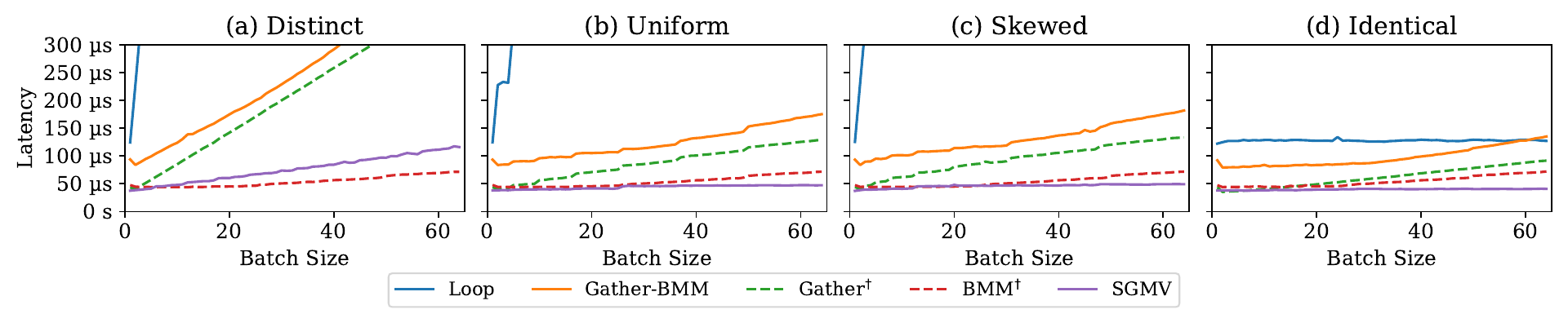}
    \caption{Microbenchmark for LoRA operator implementations. $^{\dagger}$Gather and BMM are measured separately for reference.}
    \label{fig:lora-op-impls}
\end{figure*}

\begin{figure*}[ht]
    \centering
    \includegraphics[width=\linewidth]{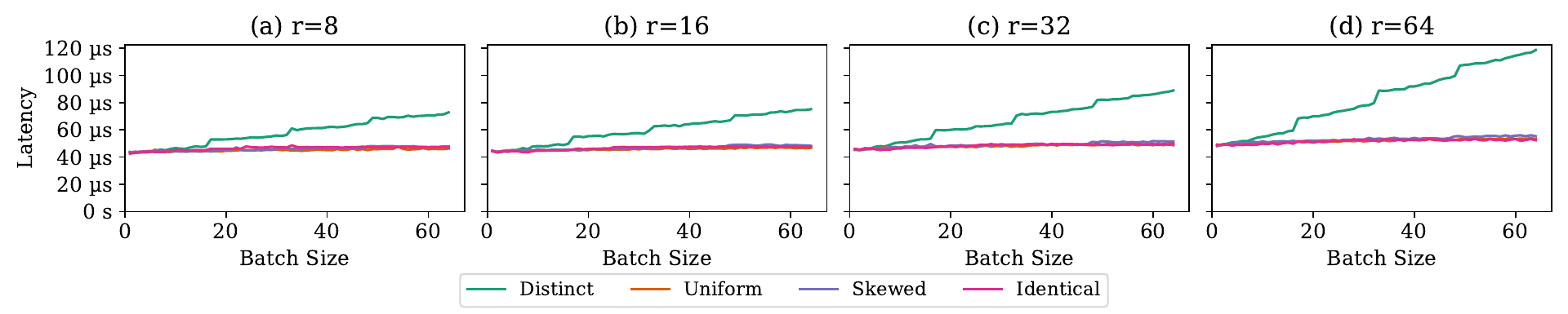}
    \caption{Microbenchmark for LoRA operator on various LoRA rank.}
    \label{fig:sgmv-varying-rank}
\end{figure*}

We implement the batched LoRA operator as two SGMV kernel launches.
We compare our SGMV-based implementation against two PyTorch-based baseline implementations.
One is a for-loop over each LoRA model.
Another is Gather-BMM.
In the gather step, we stack the weight matrices that each input needs into a single matrix.
Then, we use \verb|torch.bmm()| to perform a batched matrix multiplication on the input and the stacked matrix.
Similar to SGMV, Gather-BMM launches Gather twice and BMM twice.
Note that Gather-BMM uses much more I/O than SGMV.
Gather reads in $n \times h_i \times h_o$ elements and writes to $s_n \times h_i \times h_o$.
Then, BMM needs to read in $s_n \times h_i \times h_o$ weight elements that Gather just wrote.
In combination, Gather-BMM incurs $s_n \times h_i \times h_o \times 2$ more elements memory I/O than SGMV.

\autoref{fig:lora-op-impls} shows the latency comparison of the three implementations across four workloads on Testbed \#1. Gather and BMM are also measured separately for reference. Since BMM is data-independent, its latency is consistent across four workloads.

Our benchmark results match our analysis very well.
In the \textbf{Distinct} case, Loop behaves terribly because it runs multiple rounds of batch size 1.
Gather-BMM latency increases fast due to the slowdown of Gather.
SGMV latency increases gradually as well, from 37\textmu{}s to 116\textmu{}s, because batching does not change arithmetic intensity.
The \textbf{Uniform} case and the \textbf{Skewed} case are similar to the \textbf{Distinct} case.
Gather-BMM performs slightly better than the \textbf{Distinct} case since there are fewer matrices to read.
SGMV latency increases only marginally, from 37\textmu{}s to 46\textmu{}s, as a combination of both effects: increasing degree of parallelism and increasing arithmetic intensity.
In the \textbf{Identical} case, all implementations have the same semantics: BMM.
We can, therefore, infer that SGMV implements BMM more efficiently than \verb|torch.bmm()| in the case of LoRA.
SGMV latency remains almost constant, from 37\textmu{}s to 40\textmu{}s.

Overall, SGMV significantly outperforms baseline implementations regardless of workloads.

We also run the microbenchmark of different LoRA ranks on Testbed \#1.
\autoref{fig:sgmv-varying-rank} shows the latency for LoRA rank 8, 16, 32, and 64.
In the \textbf{Distinct} case, the latency gradually increases.
The latency of a single request batch is around 42\textmu{}s for all four ranks, while batch size 64 goes up to 72\textmu{}s, 75\textmu{}s, 89\textmu{}s, and 118\textmu{}s, respectively.
When the workload exists weight sharing (\textbf{Uniform}, \textbf{Skewed}, and \textbf{Identical}), the latency remains almost the same across batch size 1 to 64, at around 42\textmu{}s to 45\textmu{}s.

\paragraph{Transformer layer benchmark}

\begin{figure*}[ht]
    \centering
    \includegraphics[width=\linewidth]{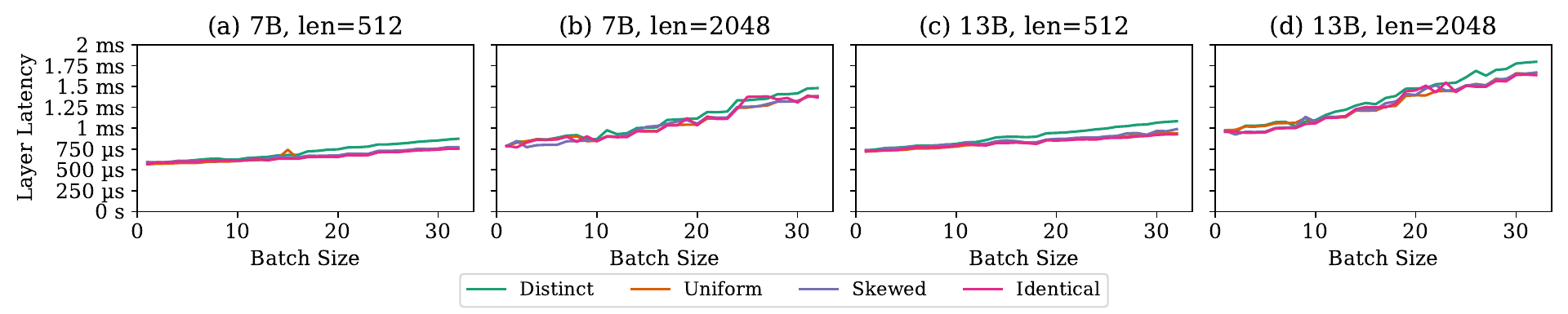}
    \caption{Transformer Layer Benchmark.}
    \label{fig:layer-latency-varying-lora-pop}
\end{figure*}

Next, we evaluate the transformer layer performance after incorporating the LoRA operator.
Since the LLM is roughly a stack of transformer layers, the layer performance determines the overall model performance.
We run the layer benchmark on Testbed \#1 based on the 7B and 13B model configurations and sequence lengths of 512 and 2048.
\autoref{fig:layer-latency-varying-lora-pop} plots the layer latency.
When the sequence length is shorter, the batching effect is stronger.
The latency only increases by 72\% when batch size increases from 1 to 32 when the sequence length is 512.
When the sequence is longer, self-attention takes longer time, which reduces the layer-wise batching effect.

In contrast to the kernel microbenchmark, notice that the layer latency is roughly the same across different workloads.
This is because the computation time for the LoRA add-on is small compared to the backbone dense projection and the self-attention.
\emph{This LoRA-model-agnostic performance property enables us to schedule different LoRA models as if one model.}
Our scheduling algorithm can then focus on the overall throughput instead of individual LoRA model placement, which is exactly how we design \sys.

\subsection{Text generation}

Next, we study the text generation performance of \sys and baseline systems.

\paragraph{Serving 7B and 13B models on a single GPU}

\begin{figure*}[ht]
    \centering
    \includegraphics[width=\linewidth]{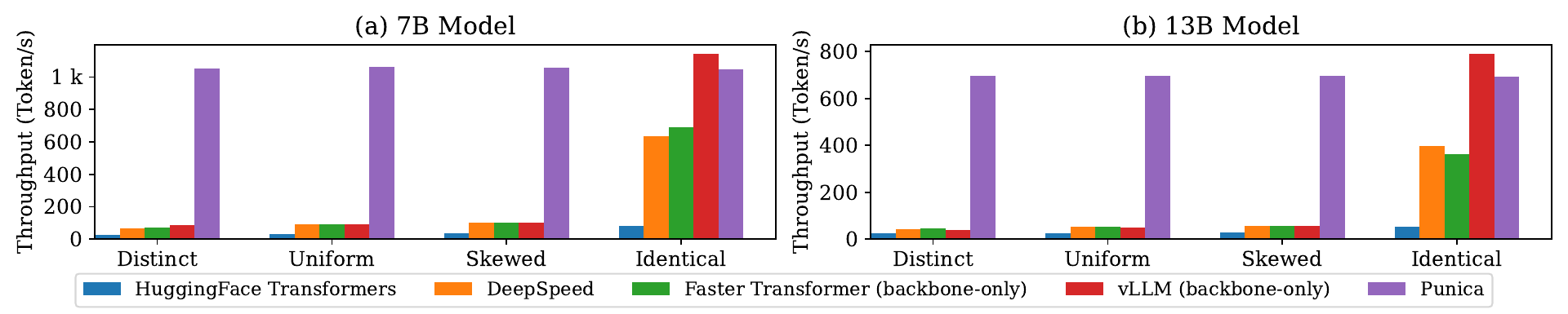}
    \caption{Single GPU text generation comparison}
    \label{fig:textgen-1gpu-with-baseline}
\end{figure*}

We evaluate text generation using \sys and baseline systems on a single GPU on Testbed \#1. The single-GPU performance serves as the base case for cluster-wide deployment.
We generate 1000 requests (generating around 101k tokens) and restrict each system to batch in a first-come-first-serve manner. The max batch size is set to 32 for all systems. \sys can batch across different LoRA models, and baseline systems can only batch requests for the same LoRA models.

\autoref{fig:textgen-1gpu-with-baseline} (a) and (b) show the results on the 7B model and the 13B model, respectively. 
\sys consistently delivers high throughput regardless of workloads.
\sys achieves 1044 tok/s and 693 tok/s on the 7B and the 13B models, respectively.
Although most baselines can achieve relatively high throughput in the \textbf{Identical} case, their performance deteriorates when there are multiple LoRA models.

In the \textbf{Distinct} case, all baseline systems run with a batch size of 1, and thus, the throughput is low.
In the \textbf{Uniform} and the \textbf{Skewed} cases, most batches for the baseline systems have extremely small batch sizes (1--3), which explains the low performance.
\sys is able to batch different LoRA models in one batch and, therefore, can run with a batch size of 32 consistently.

With only one LoRA model, all systems can run with a batch size of 32. Thus, all except for the HuggingFace Transformer can deliver high throughput. HuggingFace Transformer's low performance is due to its lack of critical CUDA kernel optimizations, including FlashAttention~\cite{FlashAttention}.
In the \textbf{Identical} case, both vLLM and \sys outperform other systems because the two systems' KvCache layout allows continuous batching. In contrast, other systems have to wait for the longest sequence in the batch to finish.
vLLM's throughput is slightly higher than \sys (at 1140 tok/s and 789 tok/s, respectively) because we run vLLM backbone-only.

\paragraph{Serving 70B models with tensor parallelism}

\begin{figure}[ht]
    \centering
    \includegraphics[width=\linewidth]{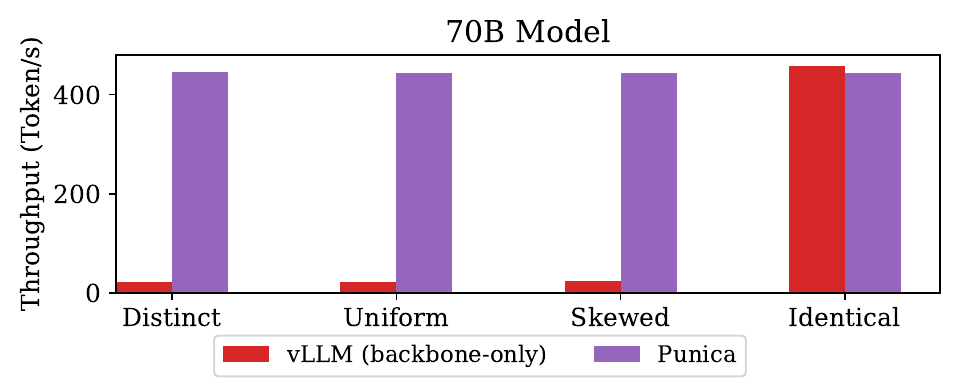}
    \caption{70B model text generation comparison.}
    \label{fig:textgen-70b}
\end{figure}

We run the 70B model on \sys with Megatron's tensor parallel scheme~\cite{megatron-lm, megatron-lm-pipeline} on 8 GPUs in Testbed \#2. We compare \sys and vLLM. vLLM also uses the same Megatron's tensor parallel scheme.

\autoref{fig:textgen-70b} shows similar trends as our results in serving 7B and 13B models.
In the presence of multiple LoRA models, vLLM's throughput is around 21 to 25 tok/s,
whereas when serving the backbone, vLLM can achieve 457 tok/s due to the large batch size.
For the \textbf{Identical} case, \sys and vLLM achieve the same performance because their parallel schemes are the same. However, \sys can consistently deliver 441 to 446 tok/s throughput regardless of LoRA popularity distribution, significantly outperforming vLLM for serving multiple LoRA models.

\subsection{Cluster deployment}

\begin{figure}[ht]
    \centering
    \includegraphics[width=\linewidth]{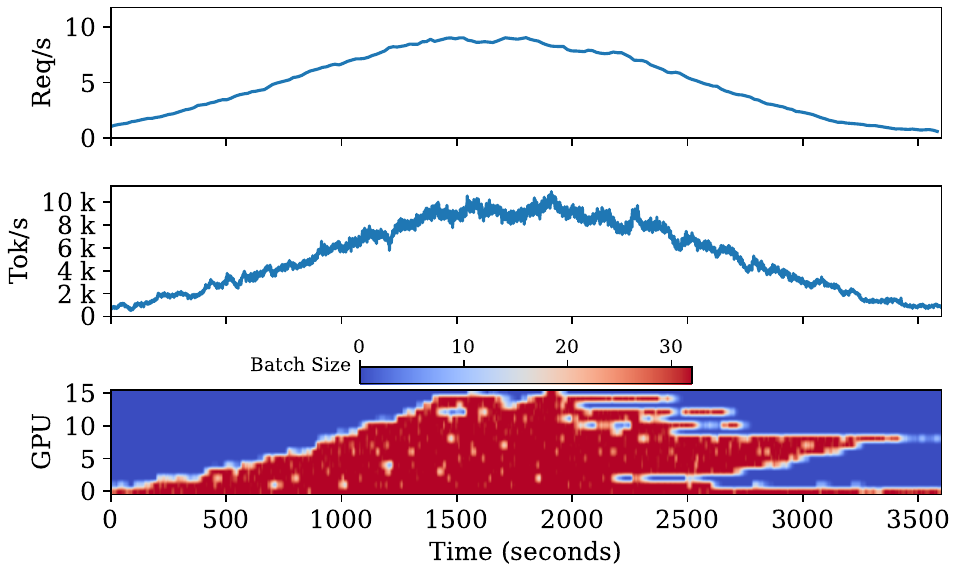}
    \caption{Cluster deployment.}
    \label{fig:cluster-deploy}
\end{figure}

We evaluate \sys on 16 GPUs in Testbed \#2. The load varies as follows: In the macro view, the request rate of the workload gradually increases and then gradually decreases. In the micro view, gaps between request arrival time follow an exponential distribution, and the arrival process follows a Poisson distribution. LoRA model popularity follows Zipf-1.5 distribution (same as our \textbf{Skewed} workload). The duration of the experiment is one hour. The model size is 7B in this experiment.

\sys is able to consolidate GPU usage while delivering high throughput. \autoref{fig:cluster-deploy}'s upper panel shows the request rate over time. The middle panel shows text generation throughput in terms of tokens per second.
The lower panel shows the batch size of each GPU across time. GPUs usually run with the maximum batch size when they are not idle because our schedule algorithm prioritizes large batch sizes. Occasionally, a GPU runs with a smaller batch size because it runs out of KvCache space and migrates out a few requests to other GPUs. When a GPU becomes idle (batch size = 0), it is likely that it stays idle, which can then be released to the cloud provider if necessary.

%% file: tex/90_related.tex
\section{Related Work}
\paragraph{LLM inference optimization.}
A series of recent work has focused on optimizing LLM inference.
Orca~\cite{orca} proposes batching transformer-based text generation by splitting concatenated batch input at the self-attention operation.
vLLM~\cite{PagedAttention} further reduces the memory fragmentation of KvCache by borrowing the idea of virtual pages in operating systems.
FlashAttention~\cite{FlashAttention} provides an optimized implementation of self-attention operation by reducing data movement via block-wise computation.
\sys already integrates them.
On the other hand, FlexGen~\cite{flexgen} designed an efficient swapping schedule to maximize throughput on a single GPU while sacrificing latency. Speculative Decoding ~\cite{speculative-decode, specinfer, medusa} increases the operational intensity of auto-regressive models by using a lightweight ``draft'' model to propose candidates for the next $k$ tokens and verifying these $k$ tokens in parallel with large models. \sys is orthogonal to these optimizations.

\paragraph{Multi-model inference serving}

A substantial body of work has been proposed for serving ML models on a GPU cluster. Clipper~\cite{clipper} is one of the earliest systems to optimize both throughput and latency in a GPU cluster. It is followed up by a series of systems~\cite{clockwork}. However, they are mainly designed to serve smaller CNN models. 
One key difference is that serving CNN models is stateless whereas LLM serving needs to persist the KvCache.
The state introduces an affinity that asks for a different system design.
For example, Symphony~\cite{symphony} uses a non-work-conserving scheduler but \sys runs batches on a GPU back-to-back due to the KvCache affinity.
Although Nexus~\cite{nexus} supports prefix sharing of different models, they offer limited support and optimization faced with LLMs and fine-grained sharing patterns as we witnessed in LoRA.

PetS~\cite{PetS} batches requests to different adapters (e.g., Adapters~\cite{Adapters}, MaskBert~\cite{MaskBert}, Diff-Pruning~\cite{Diff-Pruning}, Bitfit~\cite{BitFit}) of a LLM on a single GPU.
It allows GPU memory sharing of the pre-trained model for different downstream tasks, however, it does not enable multiple different models to run concurrently.

\paragraph{Model and KvCache quantization/compression}
A substantial amount of work has been proposed to reduce the memory footprint of model weights, activations and KvCache by quantization~\cite{gptq,smoothquant,olive,awq,flexgen}. Model quantization saves more headroom for KvCache, hence enabling \sys to serve requests of longer sequences without migration. In addition, KvCache quantization ~\cite{flexgen} and compression ~\cite{dejavu, scissorhands, heave-hitter, streamllm} further reduces the memory I/O of the KvCache, through which inference latency can be reduced, as self-attention is bounded by GPU memory bandwidth~\cite{FlashAttention}. QLoRA ~\cite{qlora} proposes to fine-tune LoRA by storing LoRA weights/gradients in high-precision formats such as fp16 while keeping the original weight in quantized formats to save memory footprint during fine-tuning. Quantization reduces self-attention latency, which makes the high efficiency of \sys's LoRA kernel even more important.

%% file: tex/99_conclusion.tex
\section{Conclusion}
Low-rank adaptation (LoRA) has become an important fine-tuning method to adapt pre-trained models to specific domains.
We present \sys, a system to serve multiple LoRA models in a shared GPU cluster. \sys's design is centered around a new CUDA kernel design that allows batching of GPU operations for different LoRA models. For each GPU, \sys only requires a single copy of the underlying pre-trained model for the GPU to serve multiple, different LoRA models, significantly improving GPU efficiency in terms of both memory and computation.
Additionally, \sys's scheduler consolidates multi-tenant LoRA serving workloads in a shared GPU cluster. With a fixed-sized GPU cluster, our evaluations show that \sys achieves 12x higher LoRA model serving throughput compared to state-of-the-art LLM serving systems while only adding 2ms latency per token.